\documentclass[pr,twocolumn,%
showkeys,showpacs,nofootinbib,byrevtex]{revtex4-1}
\usepackage{srcltx}
\usepackage{graphicx}
\usepackage{epsfig,bm,dcolumn}
\usepackage{epstopdf}
\usepackage{xcolor}
\begin{document}

\title {Role of $\omega$-meson exchange in scaling of the $\gamma p\to\pi^0 p$
process from a Regge-type model with resonances}
\author{Kook-Jin Kong}
\affiliation{ Research Institute of
Basic Sciences, Korea Aerospace University, Goyang, 412-791, Korea
}
\author{Tae Keun Choi}%
\affiliation{Department of Physics,
Yonsei University, Wonju, 26493, Korea}
\author{Byung-Geel Yu}
\email[ ]{bgyu@kau.ac.kr} \affiliation{Research Institute
of Basic Sciences,
    Korea Aerospace University, Goyang, 412-791, Korea}


\begin{abstract}
The scaling of photoproduction $\gamma p\to\pi^0 p$ is
investigated in the Reggeized model with  $N^*$ and $\Delta$
resonances included to describe resonance peaks up to photon
energy $E_\gamma$= 3 GeV. Given the $t$-channel exchanges
$\rho(770)+\omega(780)+b_1(1235)+h_1(1170)$ Reggeized for the
background contribution, the resonances of the Breit-Wigner form
are introduced to agree with cross sections for total,
differential and beam asymmetry in the low energy region. The
scaled differential-cross sections $s^7{d\sigma/ dt}$ are
reproduced to agree with the recent JLab data, revealing the
production mechanism of the big bump structure around $W\approx
2.2$ GeV by the deep-dip pattern of the $\omega$ exchange that
originates from the zeros of the trajectory $\alpha_\omega(t)=0$
in the canonical phase, ${1\over2}(-1+e^{-i\pi
\alpha_{\omega}(t)})$.
\end{abstract}

\pacs{11.55.Jy, 13.75.Gx, 13.60.Le, 12.38.Lg}
\keywords{scaling,
scaled cross section, $\pi^0$ photoproduction, $\omega$-meson,
Regge model, Breit-Wigner resonance}

\maketitle

By the modern theory of strong interaction hadrons consist of
quarks and gluons and their interactions are governed by quantum
chromodynamics. Due to the quark confinement, however, hadronic
processes are observed in terms of meson and nucleon degrees of
freedom. In this respect, it is of interest  to search for
evidences for quarks and gluon degrees of freedom shown up, in
particular, in exclusive hadronic processes.

Recent experiments on pion photoproduction at JLab
\cite{zhu,bartholomy,dugger} have opened such a possibility by
measuring the  differential cross section at high energy and large
momentum transfer which is predicted to obey the scaling, i.e.,
\begin{eqnarray}
    {d\sigma\over dt}\sim F(t/s)s^{2-n}\,,
\end{eqnarray}
based on the quark-counting rule \cite{brodsky1,brodsky2}. Here
$n$ is the number of constituents (gauge boson plus the quarks)
participating in the process. For the pion photoproduction,
therefore, the measured cross sections in the CLAS g1c
collaboration are expected to show such a scaling behavior
$s^7{d\sigma/dt}\sim F(t_0/s)\sim$ constant at the fixed angle
around $\theta=90^\circ$ (or fixed $t_0$) as energy increases.

In this letter we investigate the process $\gamma p\to \pi^0 p$
with a focus on understanding the origin of the scaling at the
level of hadronic degrees of freedom. As the existing data on this
issue requires a model capable of reproducing not only  the
resonance peaks in the low energy but also the deep dip process
observed at high energy it is advantageous to work with the
reggeized model which includes nucleon resonances \cite{chiang}.
Since the Regge pole  is in essence a partial wave analytically
continued to the complex angular momentum space, it is good to
consider the nucleon resonance as the partial wave of the
Breit-Wigner (BW) form for a specific angular momentum and parity
eigenstate in the production amplitude.
%

In the neutral process where $\pi^0$ exchange is forbidden by
charge-conjugation, $\omega(780),\,\rho(770),\, h_1(1170)$, and
$b_1(1235)$ Regge poles are considered in the $t$-channel exchange
to serve as a background contribution. Then, the whole amplitude
will be composed of the $t$-channel Regge poles and the nucleon
resonance $R$ of spin-$J$ possible to the process, i.e.,

\begin{eqnarray}\label{totalamp}
{\cal M}={\cal M}_{Regge}+\sum_{R=\Delta}{\cal
M}_{R}^J+\sum_{R=N^*}{\cal M}_{R}^J\,.
\end{eqnarray}

We recall that the $\omega$ exchange with the phase ${1\over
2}(-1+e^{-i\pi \alpha_\omega})$ and the $\rho$ with $e^{-i\pi
\alpha_\rho}$ are favored to describe the deep dip in the
differential cross section in the reggeized model \cite{glv}. For
a better agreement with the beam asymmetry $\Sigma$ at high energy
the exchanges of $b_1$ and the $h_1$ are further introduced with
the constant phase taken \cite{bgyu}. In Table \ref{tb1} the
coupling constants and the Regge trajectories are summarized with
the detailed expression for the ${\cal M}_{Regge}$ given in Ref.
\cite{bgyu}.

\begin{table}[t]
    \caption{Coupling constants and trajectories for Regge poles. $g_{\gamma\pi V}$ is
    given in unit of GeV$^{-1}$.}
    \begin{tabular}{cccccc}
        meson & trajectory & phase & $g_{\gamma\pi V}$ & $g^v_{VNN}$ & $g^t_{VNN}$  \\
        \hline \hline
        $\rho$ & $0.8t+0.55$ & $e^{-i\pi \alpha_{\rho}}$ & 0.255 & 2.6 & 16.12 \\%
        $\omega$ &  $0.9t+0.44$ & ${-1+e^{-i\pi \alpha_{\omega}}\over2}$ & 0.723 & 15.6 & 0 \\%
        $b_1$ &  $0.7(t-m_{b_1}^2)+1$ & 1 & 0.189 & 0 & $-14$ \\
        $h_1$ &  $0.7(t-m_{h_1}^2)+1$ & 1 & 0.405 & 0 & $-14$ \\%
        \hline
    \end{tabular}\label{tb1}
\end{table}

We now consider the resonances that are ranked as  the four-star
in Particle Data Group (PDG) for definiteness sake. For each
resonance the partial wave is described by the helicity amplitude
of Ref. \cite{walker} with the definition and the sign convention
of the electric and magnetic multipoles further from Refs.
\cite{burkert,tiator} in relation with the Chew, Goldberger,Low
and Nambu (CGLN) amplitude. Then, the resonance $R$ of spin-$J$ is
written as
\begin{eqnarray}\label{bw}
{\cal M}_{R}^J&=&\,\sum_{i=1}^4 H_i^J(W,\theta),
\end{eqnarray}
where the angular dependence of the amplitude $H_i^J$ is expressed
in the helicity formulation as in Ref. \cite{walker} with its
energy-dependence given by the electric and magnetic multipoles
\cite{thom},
\begin{eqnarray}\label{bw1}
E_{l\pm} &=& {\beta_E \over \sqrt{q_Rk_Rj_\gamma(j_\gamma+1)}}
\frac{ M_R\sqrt{\Gamma_{R\gamma N} \Gamma_{R\pi
N}}}{M^2_R-s-iM_R\Gamma}
\,e^{-d\epsilon^2_R} \,,\\
M_{l\pm}&=& {\beta_M \over \sqrt{q_Rk_Rj_\gamma(j_\gamma+1)}}
\frac{ M_R\sqrt{\Gamma_{R\gamma N} \Gamma_{R\pi
N}}}{M^2_R-s-iM_R\Gamma}\,e^{-d\epsilon^2_R} \,,\label{bw2}
\end{eqnarray}
respectively.

Note that the electromagnetic multipoles in Eqs. (\ref{bw1}) and
(\ref{bw2}) are different from those in Ref. \cite{thom} in that
the decay widths of resonance $R\to\gamma N$ and $R\to\pi N$
reported in the PDG are employed rather than $\Gamma_E$,
$\Gamma_M$ and $\Gamma_{l\pm}$ with cutoff parameters in Ref.
\cite{thom} for less model-dependence.
Instead, we use the $\beta_{E}$ and $\beta_{M}$ which are to be
determined at the resonance position $W=M_R$ by using Eqs.
(\ref{bw1}) and (\ref{bw2}).  To be specific, $\beta_E(\beta_M)$
is determined by estimating $E_{l\pm}(M_{l\pm})$ and
$\Gamma_{R\gamma N}$ in Eqs. (\ref{bw1}) and (\ref{bw2}) at
$W=M_R$ from the following procedure \cite{burkert}; Given the
values of the helicity amplitudes $A_{1/2}$ and $A_{3/2}$  for
$l+$ multipole amplitudes,
\begin{eqnarray}
A_{1/2} &=&-{1 \over 2}\left[(l+2){\cal E}_{l+}+l{\cal M}_{l+}\right], \\
A_{3/2} &=&{\sqrt{l(l+2)} \over 2}\left[{\cal E}_{l+}-{\cal
M}_{l+}\right],
\end{eqnarray}
and for $(l+1)-$ multipole amplitudes,
\begin{eqnarray}
A_{1/2} &=&{1 \over 2}\left[(l+2){\cal M}_{(l+1)-}-l{\cal E}_{(l+1)-}\right], \\
A_{3/2} &=&-{\sqrt{l(l+2)} \over 2}\left[{\cal E}_{(l+1)-}+{\cal
M}_{(l+1)-}\right],
\end{eqnarray}
where the multipoles  ${\cal E}_{l\pm}({\cal M}_{l\pm})$ are given
by
\begin{eqnarray}\label{cgln1}
a{\cal E}_{l\pm}({\cal M}_{l\pm})={\rm Im}\, E_{l\pm}(M_{l\pm}),
\end{eqnarray}
the $E_{l\pm}(M_{l\pm})$ can be estimsted at $W=M_R$ with
\begin{eqnarray}
&& a=C_I\left[{1\over (2J+1)\pi} {k_R\over q_R}{M_N\over M_R}
{\beta_{\pi N}\over \Gamma}\right]^{1/2}\,,\\
&& C_{1/2}=-\sqrt{1\over 3}\ ,\ \ \ \  C_{3/2}=\sqrt{2\over 3}\ .
\end{eqnarray}
Here $\beta_{\pi N}$ is the branching ratio of the resonance to
the  $\pi N$ channel. $\Gamma$, $M_R$, $J$ and $I$ are the total
width, mass, spin, isospin of the resonance $R$. $k_R$ and $q_R$
are momenta of  photon and pion at the resonance position in the
c.m. system. $C_I$ is the Clebsch-Gordan coefficient and $M_N$ is
the nucleon mass.

With the known values for $A_{1/2}$ and $A_{3/2}$ the partial
width $R\to\gamma N$ in Eqs. (\ref{bw1}) and (\ref{bw2}) is
obtained as well from the equation,
\begin{eqnarray}
\Gamma_{R\gamma N}={k_R^2 \over
\pi}\frac{2M_N}{(2J+1)M_R}\left[|A_{1/2}|^2+|A_{3/2}|^2\right].
\end{eqnarray}

In relation with Ref. \cite{thom}, it is legitimate to write
\begin{eqnarray}\label{bw9}
\Gamma_{E(M)} \Gamma_{l\pm}=\beta^2_{E(M)} \Gamma_{R\gamma N}
\Gamma_{R\pi N}\ .
\end{eqnarray}
Then, the $\beta_{E}$($\beta_{M}$) estimated as discussed above,
thus, provides the information about the case of coupling of the
electric (magnetic) photon in the model of Ref. \cite{thom}. For a
better description of the resonance we now introduce the cutoff
factor $e^{-d\epsilon^2_R}$ to adjust the range of the resonance
by a dimensionless parameter $d$ together with the
$\epsilon_R={M_R^2-s \over M_R\Gamma}$ \cite{lennox}. Therefore,
there is no parameter, in essence, except for the $d$ within the
present framework where $M_R$, $\Gamma$, $\beta_{\pi N}$, and
$A_{1/2}$ and $A_{3/2}$ are to be taken from PDG.

\begin{figure}[t]
\includegraphics[width=8.6cm]{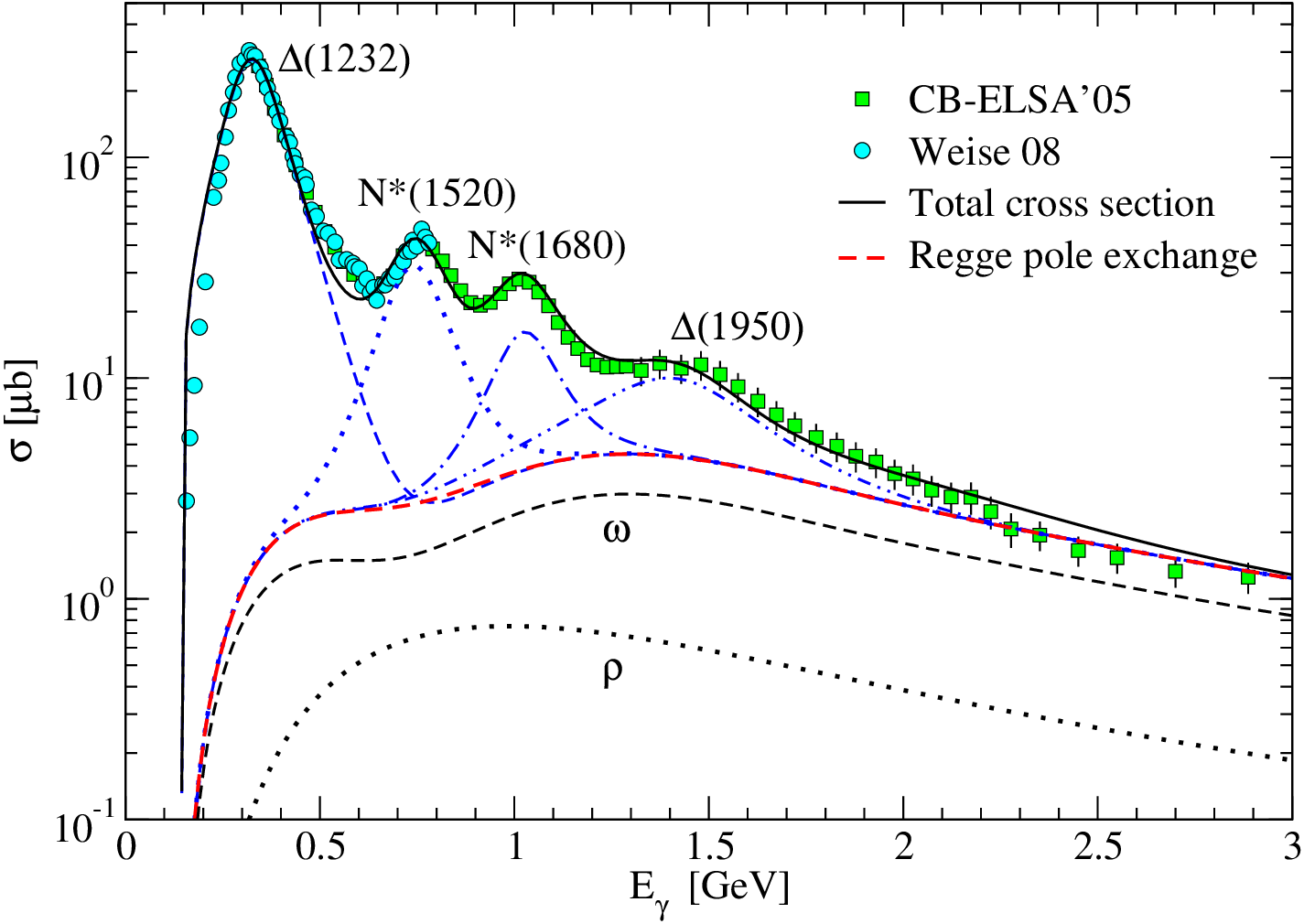}
\caption{(Color online) Total cross section for $\gamma\, p \to
\pi^0\, p$ from threshold to $E_\gamma$=3 GeV.  Red dashed line is
the background from $\rho+\omega+b_1+h_1$ Regge pole exchanges.
Black solid line results from the background plus the resonances.
Four resonances $\Delta(1232)$ (dashed), $N(1520)$ (dotted),
$N(1680)$ (dash-dotted), and $\Delta(1950)$  (dash-dot-dotted) are
illustrated to represent the sequential peaks. Data are taken from
Refs. \cite{bartholomy,wise}.} \label{fig1}
\end{figure}

\begin{figure}[t]
\vspace{0.3cm}
\includegraphics[width=8.6cm]{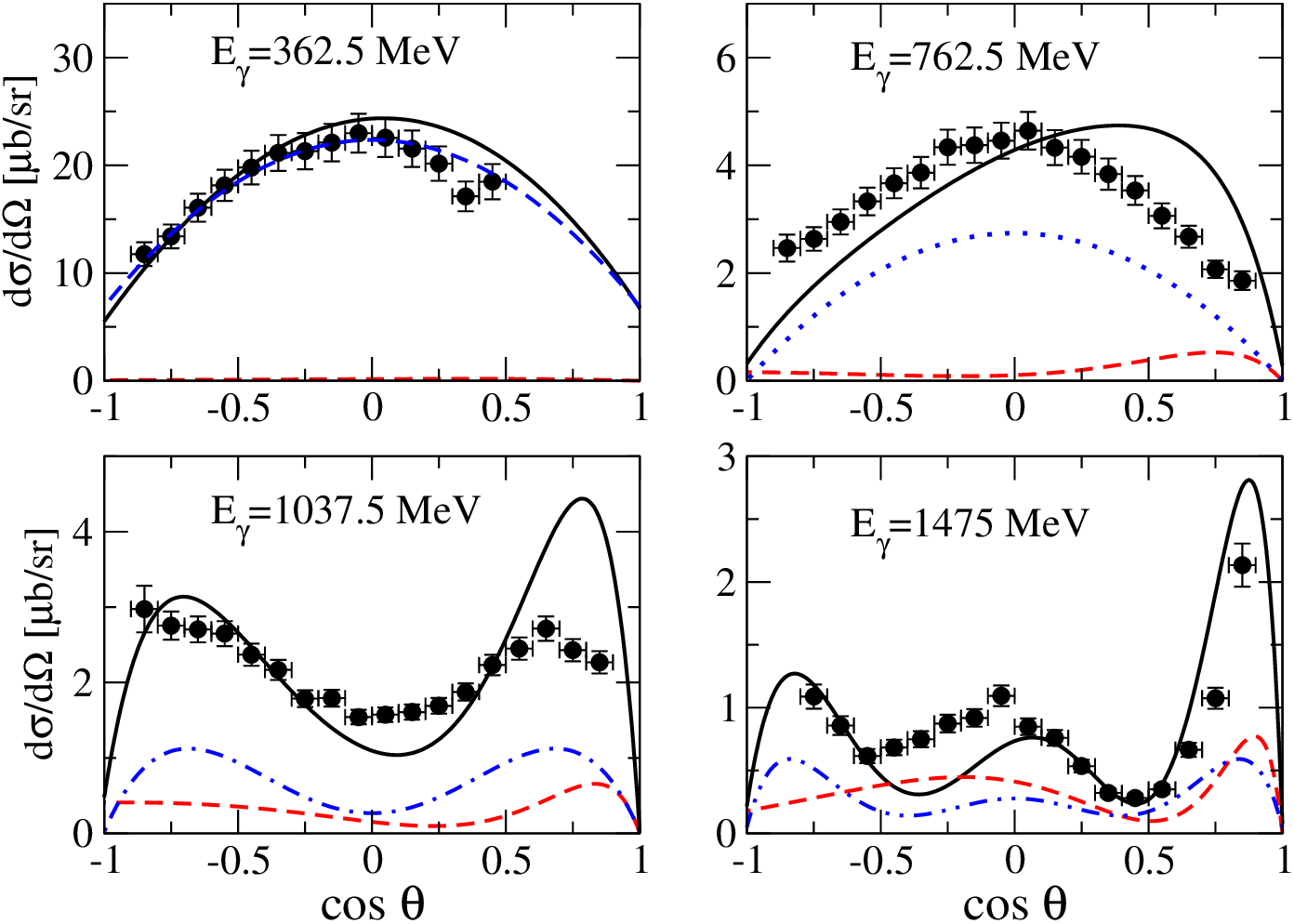}
\caption{(Color online) Differential cross sections for $\gamma\,
p \to \pi^0\, p$ in four energy bins $E_\gamma$=362.5
($W$=1249.2), $762.5\ (1520.3)$, $1037.5\ (1681.5)$ and $1475\
(1910)$ MeV. Data are taken from Ref. \cite{bartholomy}. Notations
    for the curves are the same as in Fig. \ref{fig1}.} \label{fig2}
\end{figure}

Before proceeding we wish to give a remark on the possibility of
double-counting by the duality of the $t$-channel Regge poles to
the $s$-channel resonances. The double counting should be
considered in case when the contributions of the $t$-channel
exchanges amount to an average of the resonance peaks in
differential and, as a result, in total cross section
\cite{lennox}. It is expected that the negligence of it does not
cause a serious problem in the neutral pion case, however, because
the contributions of the $t$-channel exchanges are not so much as
those of the $s$-channel resonances to be taken an average, as can
be seen in Figs. \ref{fig1} and \ref{fig2}. Henceforth, we neglect
the double-counting between the resonances and the Regge poles in
the present calculation.

With the physical constants listed in table \ref{tb2} which are
within the range of PDG values, we present the total cross section
in Fig. \ref{fig1} up to $E_\gamma=3$ GeV where the contributions
of the resonances, $\Delta(1232)$, $N(1520)$, $N(1680)$, and
$\Delta(1950)$ are shown to represent the four prominent peaks,
respectively. The respective contributions of the $\omega$ and
$\rho$ Regge pole exchanges are illustrated by the dashed and
dotted lines. In the Regge realm beyond $E_\gamma$=3 GeV, of
course, the cross section coincides with that from the pure Regge
pole exchanges as described in Ref. \cite{bgyu}.

\begin{table}[b]
    \caption{Physical constants for $\Delta$ and $N^*$ resonances for $
    \gamma p\to\pi^0 p$.
    Widths and masses are given in unit of MeV, $A_{1/2}$, $A_{3/2}$ are given in unit of
    GeV$^{-1/2}$.}
    \begin{tabular}{cccccccc}
        Resonance & $J^P$ & $M_R$ & $\Gamma$ & $A_{1/2}$ &  $A_{3/2}$ & $\beta_{\pi N}$  & $d$   \\
        \hline \hline
        $\Delta(1232)$& $(3/2)^+$ & 1220 & 130  & $-0.13$ & $-0.26$ & 1.0  & 0.05  \\%
        $N(1440)$ & $(1/2)^+$ & 1400 & 190  & $-0.08$ &0.0 & 0.75 & 0.25 \\%
        $N(1520)$ & $(3/2)^-$ & 1505 & 120 & $-0.007$ & 0.168 & 0.65 & 0.1 \\%
        $N(1535)$ & $(1/2)^-$ & 1515 & 150 & 0.13 & 0.0 & 0.55 & 0.25 \\%
        $\Delta(1620)$ & $(1/2)^-$ & 1590 & 120 & 0.055 & 0.0 & 0.3 & 0.25 \\%
        $N(1650)$ & $(1/2)^-$ & 1670 & 120 & 0.065 & 0.0 & 0.7 & 0.1 \\%
        $N(1675)$ & $(5/2)^-$ & 1660 & 135 & 0.01 & 0.015 & 0.35 & 0.2 \\%
        $N(1680)$ & $(5/2)^+$ & 1675 & 110 & $-0.01$ & 0.10 & 0.675 & 0.1 \\%
        $\Delta(1700)$ & $(3/2)^-$ & 1675 & 250 & $0.12$ & 0.12 & 0.15 & 0.25 \\%
        $N(1720)$ & $(3/2)^+$ & 1680 & 250 & 0.07 & 0.15 & 0.1 & 0.1 \\%
        $\Delta(1905)$ & $(5/2)^+$ & 1900 & 280 & $0.025$ & $-0.04$ & 0.1 & 0.2 \\%
        $\Delta(1910)$ & $(1/2)^+$ & 1885 & 220  & $0.06$ &0.0 & 0.3 & 0.25 \\%
        $\Delta(1950)$ & $(7/2)^+$ & 1890 & 240 & $-0.02$ & $-0.10$ & 0.35 & 0.1 \\%
        $N(2190)$ & $(7/2)^-$ & 2150 & 450 & $-0.065$ & $ 0.035$ & 0.1 & 0.2 \\%
        $N(2220)$ & $(9/2)^+$ & 2250 & 400 & $0.01$ & $ 0.01$ & 0.2 & 0.2 \\%
        $N(2250)$ & $(9/2)^-$ & 2275 & 500 & $0.01$ & $ 0.01$ & 0.1 & 0.2 \\%
        \hline
    \end{tabular}\label{tb2}
\end{table}

\begin{figure}[t]
\includegraphics[width=8.6cm]{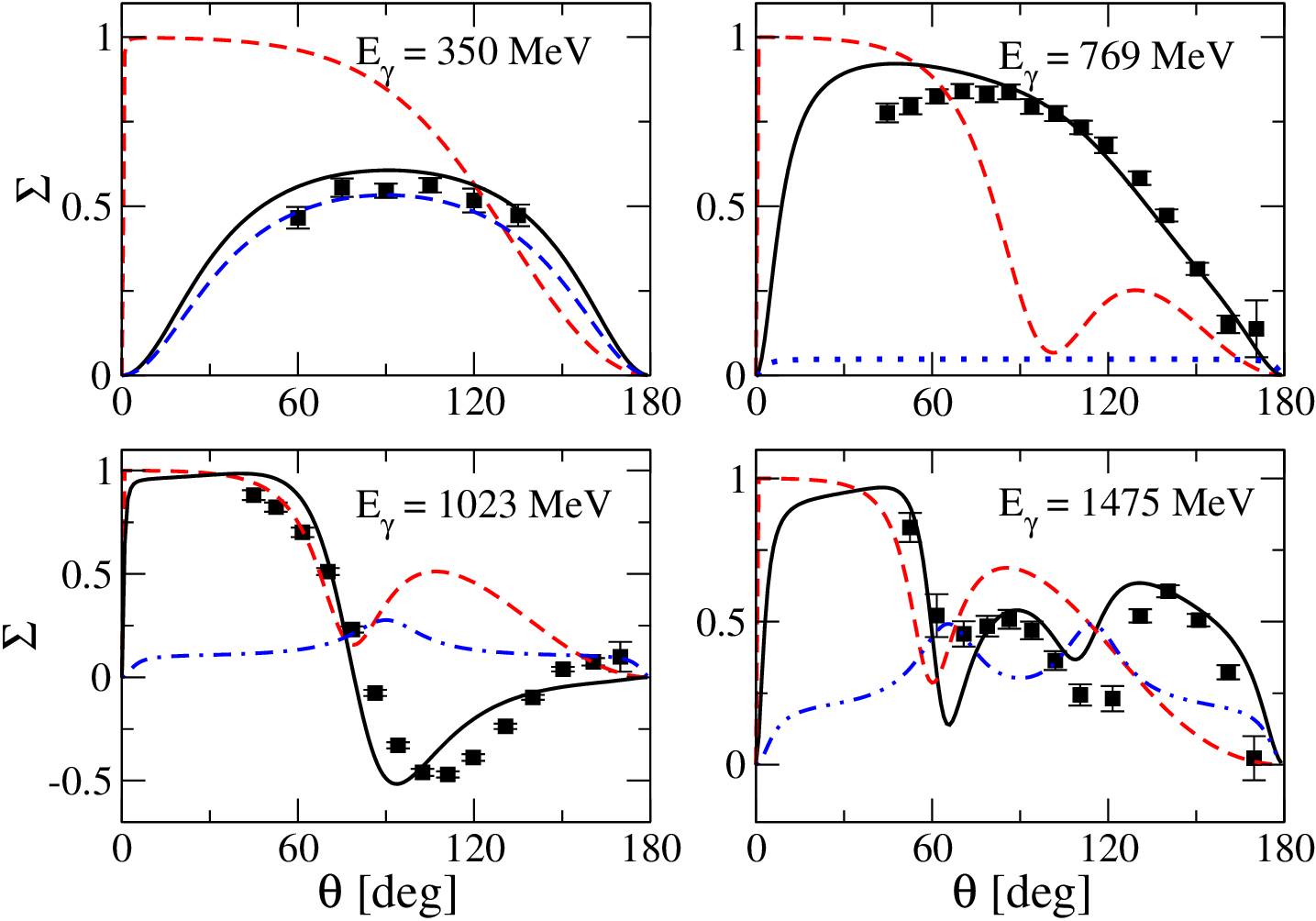}
\caption{Angular distribution of beam asymmetry $\Sigma$ for
$\gamma\, p \to \pi^0\,p$ in four energy bins $E_\gamma$=350
($W$=1240), $769\ (1524)$, $1023\ (1673)$ and $1475\ (1910)$ MeV.
Data are taken from Refs. \cite{bartalini,belyaev}. Notations for
the curves are the same as in Fig. \ref{fig1}.} \label{fig3}
\end{figure}

Figures \ref{fig2} and \ref{fig3} show the differential cross
sections and the beam asymmetries reproduced at four energy bins
each of which corresponds to the region around $\Delta(1232)$,
$N(1520)$, $N(1680)$, and $\Delta(1950)$, respectively. The solid
line is the total sum of the resonances and  the $t$-channel Regge
poles. The background contribution depicted by the red dashed line
results from the $t$-channel Regge pole exchanges. The blue
dashed, dotted, dashed-dot and dash-dot-dotted lines are from the
$\Delta(1232)$, $N(1520)$, $N(1680)$, and $\Delta(1950)$ in order.
These results are obtained through an optimal compromise between
cross sections for total, differential, and beam asymmetry to
agree with experimental data.
Near the first resonance peak $W\approx$ 1232 MeV, it is enough to
consider $\Delta(1232)$ to describe the differential cross section
and beam asymmetry. In the second resonance region mainly due to
$N$(1520), the resonances $N(1440)$ and $N(1535)$ are added   to
agree with the beam asymmetry $\Sigma$. To reproduce the
differential cross section at the third peak due to $N$(1680), the
resonance $\Delta(1700)$  together with $N(1650)$ play the role to
fit to $\Sigma$. Around the fourth peak, apart from $\Delta(1950)$
which is a most representative one to cover this region,
$\Delta(1910)$ is considered to give a further contribution to
$\Sigma$.
\begin{figure}[t]
\centering
\includegraphics[width=8.6cm]{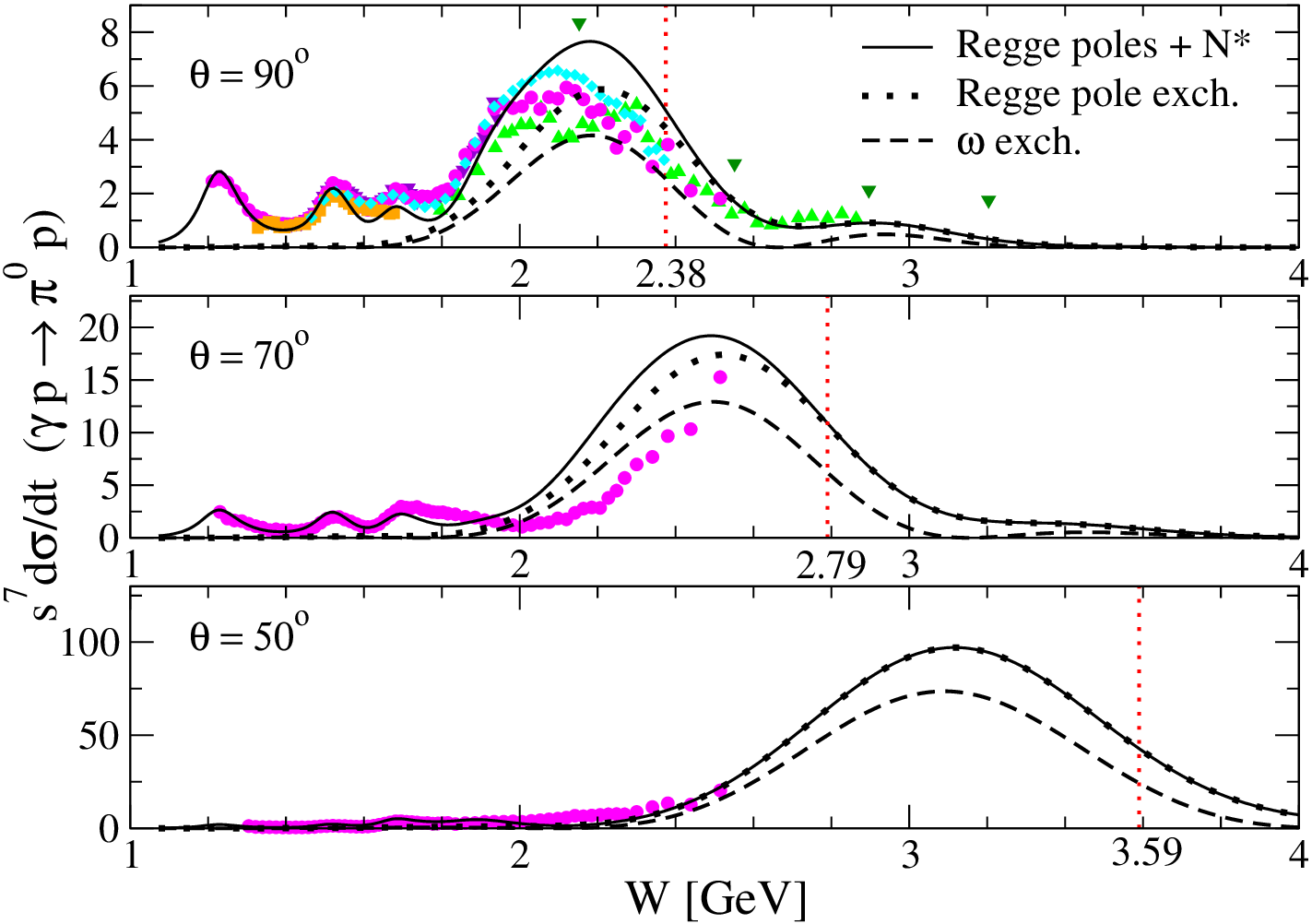}
    \caption{(Color online) Scaled cross sections
    $s^7{d\sigma\over dt}(10^7\mbox{GeV}^{14}\mbox{nb}/\mbox{GeV}^2)$
    for $\gamma\,p$$ \to $$\pi^0\, p$ at $\theta=90^\circ$, $\theta=70^\circ$,
    and $\theta=50^\circ$
    as a function of total energy W. The scaled cross section agrees
    with data (green triangle-up) up to
    $W\approx 3$ GeV. $N^*$ contributions vanish at $\theta=50^\circ$.
    For the given angle $\theta$ the red dotted vertical line indicates
    the point of $W$ corresponding to the upper
    limit of $t=-2$ GeV$^2$  maximally allowed before
    saturation in Eq. (\ref{saturation}).
    Data are taken from Refs.
    \cite{shupe,alverez,bartholomy,dugger,bartalini,imanishi}.}\label{fig4}
\end{figure}

Let us now discuss the the feature of the differential cross
section scaled by $s^7$ based on the framework we have established
in the low energy region as well as at high energy. Figure
\ref{fig4} shows three scaled cross sections the first of which
reproduces existing data at the angle $\theta=90^\circ$, while the
other two correspond to the cases  at $\theta=70^\circ$ and
50$^\circ$, respectively. It is certain that the oscillatory
behavior below $W\approx$ 2 GeV is due to the resonances
$\Delta(1232)$, $N(1520)$, and $N(1680)$ in order. Within the
Regge framework where the production mechanism of the $\pi^0$
process is dominated by the $\omega$ exchange, the bump structure
around $W\approx 2.2$ GeV, in essence, results from the deep-dip
pattern that arises from the nonsense wrong signature zero (NWSZ)
of the $\omega$ trajectory $\alpha_\omega(t)=0$ \cite{bgyu,glv}.
This in fact leads to  the vanishing of the amplitude at $W\approx
1.58$ GeV at the angle $\theta=90^\circ$, as shown in Fig.
\ref{fig4}. While the cross section is suppressed in the low
energy region, but amplified at high energy by the $s^7$-power,
the shift of the bump to the position $W\approx$ 2.5 GeV at
$\theta=70^\circ$, and to 3.1 GeV at $\theta=50^\circ$ in Fig.
\ref{fig4} can, thus, be understood as the result of a sequential
shift of the NWSZ to $W\approx 1.75$ GeV, and to 2.11 GeV
accompanying with the diffractive pattern according to  the change
of the angle. The resonance structures apparent in the scaled
cross section at $\theta=90^\circ$ vanish by degrees without
affecting the structure of the bump as the angle becomes forward
directional. This confirms our understanding of the production
mechanism of the bump structure by the diffraction of $\omega$
within the present framework.

It is worth noting that the $\rho$N cut plays a role in the scaled
cross section for the cases of $\pi^\pm$ processes \cite{laget}.
In the $\pi^0$ case  the $\omega$N cut will be the corresponding
one consisting of the amplitude $T_{\gamma p\to\omega p}T_{\omega
p\to\pi^0 p}\ $. It may work in the present process as well with
the $b_1$-pole in the latter amplitude. Nevertheless we leave the
estimate of the $\omega$N cut behind because it is beyond the
scope of the present work.

\begin{figure}[t]
    \centering
\includegraphics[width=7.8cm]{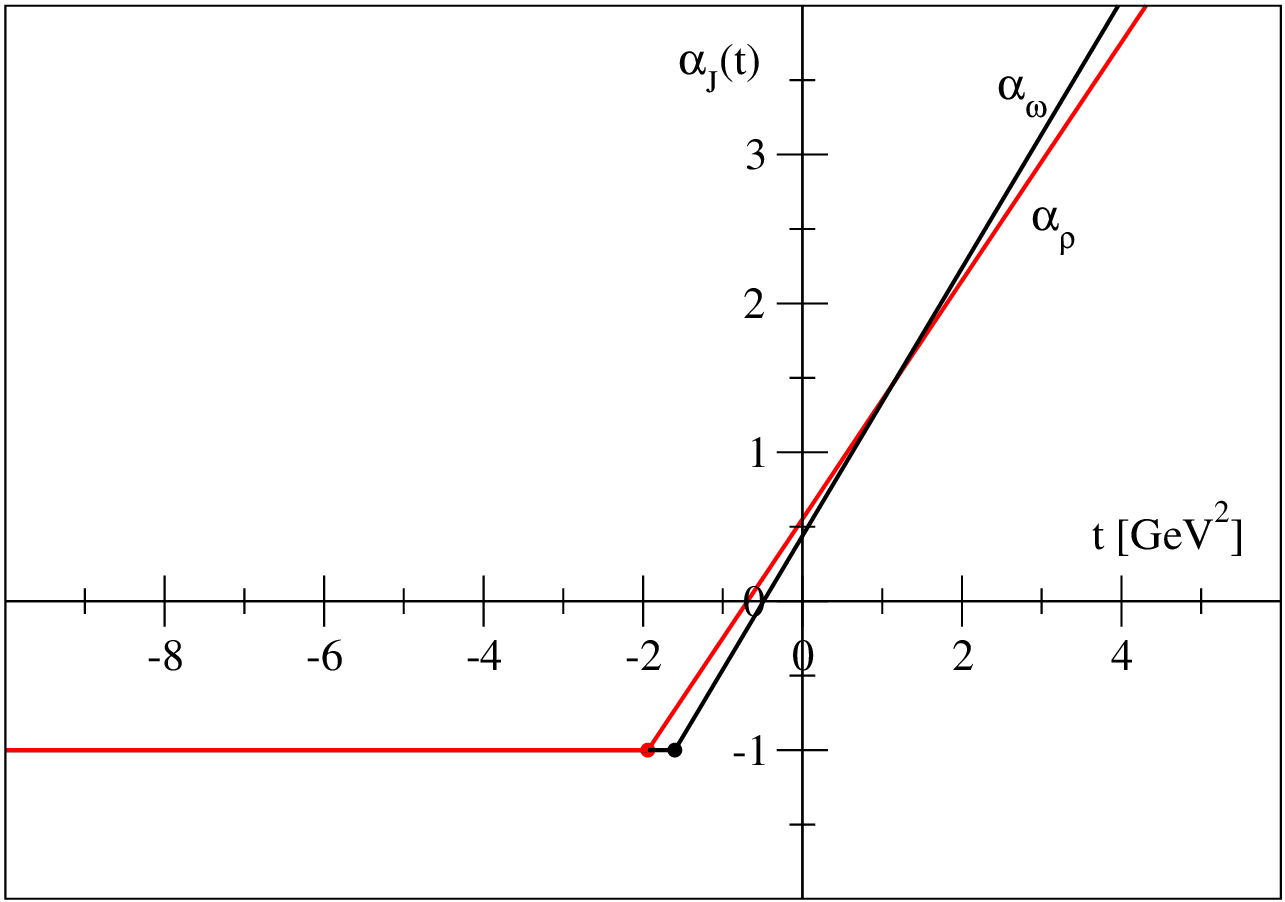}
    \caption{(Color online) Saturation of the trajectory
    $\alpha(t)=\alpha_0+\alpha' t$.
The maximum ranges of $t$ limited by $\alpha_i(t)>-1$ are
$t>-1.94$ for $\rho$, $t>-1.6$ for $\omega$, $t>-1.34$ for $b_1$,
and $t>-1.49$ GeV$^2$ for $h_1$, respectively. }
    \label{fig5}
\end{figure}

Before closing the discussion of the scaling, we must ask
ourselves whether such an analysis of the present model is valid
for the large angle, $\theta\approx 90^\circ$ in Fig. \ref{fig4}.
The Regge theory is, in general, applied in the kinematical
region, $s\sim$ large and $t<0$. Furthermore, in the large $-t$
limit the Regge trajectory is found to saturate \cite{brodsky3}
\begin{eqnarray}\label{saturation}
\lim_{t\to-\infty}\alpha(t)\to -1\,,
\end{eqnarray}
as shown in Fig. \ref{fig5}. This puts a limit on the validity of
the linear trajectory $\alpha(t)=\alpha_0+\alpha't$ in the Regge
pole. For practical interest, here, we examine how far the range
of energy $W$ in the scaling could be accepted in accord with the
point of $t$ maximally allowed before saturation as in Eq.
(\ref{saturation}), i.e., at $-t=2$ GeV$^2$ from Fig. \ref{fig5}.
The $t$-channel momentum transfer,
\begin{eqnarray}
t&&=m_\pi^2-2k\left[\sqrt{q^2+m^2_\pi}-q\cos\theta\right],
\end{eqnarray}
with this value at $\theta=90^\circ$ leads to $W$=2.38 GeV, which
could be an upper bound for the validity of the present analysis.

In summary, we have investigated the $\gamma p\to \pi^0 p$ process
with a view to understanding the bump structure not expected from
nucleon resonances but observed in the scaled cross section
$s^7d\sigma/dt$. It is pointed out that the big bump structure
originates substantially from the dip-generation mechanism of the
$\omega$-meson exchange at the NWSZ in the scaled cross section
within the present framework. For a reliable discussion on this
issue we have proved an agreement with existing data on
differential, total and the beam asymmetry, in particular, in the
low energy region by
includig nucleon resonances basically without fit-parameters. Such
resonance peaks in the scaled cross sections are reproduced well
enough to confirm the validity of the scaling discussed in this
work. The present approach will provide us a useful framework to
analyze photoproduction data from threshold up to the energy
region, coming of 12 GeV-upgrade at JLab for the forward angle,
and to $3$ GeV without the $\omega$N cut for the intermediate
angle where the scaled cross section is analyzed.

This work was supported by the grant NRF-2013R1A1A2010504 from
National Research Foundation (NRF) of Korea.


\end{document}